\documentclass[aps,prl,showpacs,twocolumn,superscriptaddress]{revtex4}
\usepackage[latin1]{inputenc}
\usepackage{graphicx}
\usepackage{color}
\begin{document}

%\preprint{}

%\title{Pressure dependence of the atomic displacements in the
%PbMg$_{1/3}$Ta$_{2/3}$O$_3$ relaxor ferroelectric}
\title{Anomalous pressure dependence of the atomic displacements in the
relaxor ferroelectric PbMg$_{1/3}$Ta$_{2/3}$O$_3$}

% Please use the appropriate macro for each type of information


\author{S.N. Gvasaliya}
\email[]{severian.gvasaliya@psi.ch}
%\thanks{}
\altaffiliation{On leave from Ioffe Physical Technical Institute,
26 Politekhnicheskaya, 194021, St. Petersburg, Russia}
\affiliation{Laboratory for Neutron Scattering ETHZ \& Paul Scherrer
Institut, CH-5232 Villigen PSI, Switzerland}

\author{V. Pomjakushin}
%\thanks{}
\affiliation{Laboratory for Neutron Scattering ETHZ \& Paul Scherrer
Institut, CH-5232 Villigen PSI, Switzerland}

\author{B. Roessli}
%\thanks{}
\affiliation{Laboratory for Neutron Scattering ETHZ \& Paul Scherrer
Institut, CH-5232 Villigen PSI, Switzerland}

\author{Th. Str$\rm\ddot a$ssle}
%\email[]{your}
%\thanks{}

\affiliation{Laboratory for Neutron Scattering ETHZ \& Paul Scherrer
Institut, CH-5232 Villigen PSI, Switzerland}
\affiliation{Physique des Milieux Denses, IMPMC, CNRS-UMR 7590,
Universit$\rm\acute e$ Pierre et Marie Curie, F-75252 Paris, France}

\author{S. Klotz}
%\email[]{your}
%\thanks{}
\affiliation{Physique des Milieux Denses, IMPMC,
CNRS-UMR 7590, Universit$\rm\acute e$ Pierre et Marie Curie, F-75252 Paris, France}

\author{S.G. Lushnikov}
%\email[]{your}
%\thanks{}
\affiliation{Ioffe Physical Technical Institute, 26 Politekhnicheskaya,
194021, St. Petersburg, Russia}

\date{\today}

\begin{abstract}
The crystal structure of the PbMg$_{1/3}$Ta$_{2/3}$O$_3$ (PMT)
relaxor ferroelectric was studied under hydrostatic pressure up to
$\sim 7$~GPa by means of powder neutron diffraction. We find a
drastic pressure-induced decrease of the lead displacement from the
inversion centre which correlates with an increase by $\sim$~50 \%
of the anisotropy of the oxygen temperature factor. The vibrations
of the Mg/Ta are, in contrast, rather pressure insensitive. We
attribute these changes being responsible for the previously
reported pressure-induced suppression of the anomalous dielectric
permittivity and diffuse scattering in relaxor ferroelectrics.

%The pressure-induced suppression of the peak in the dielectric permittivity
%and of the diffuse scattering in relaxor ferroelectrics are attributed
%to such changes in the structure.
%Our results support the ionic
%displacements picture for the occurrence of the dielectric anomaly in
%relaxor ferroelectrics.
\end{abstract}


\maketitle

\noindent

AB$'_x$B$''_{1-x}$O$_3$ complex perovskites are model compounds for
the study of ferroelectricity in disordered crystals. In many of
these compounds ferroelectricity occurs without a well-defined
structural phase transition, and consequently they are referred as
relaxor ferroelectrics, or simply relaxors~\cite{smol1}. Relaxors
have a frequency dependent peak in the dielectric permittivity which
typically extends over hundreds of degrees Kelvin.  In addition,
many physical properties of relaxors exhibit anomalies in this
temperature range attributed to a so-called "diffuse phase
transition". Despite numerous studies, the physics of systems
exhibiting diffuse phase transitions is not well understood yet.

In the course of recent
studies on the typical relaxor ferroelectrics
PbMg$_{1/3}$Nb$_{2/3}$O$_3$ (PMN) and PbMg$_{1/3}$Ta$_{2/3}$O$_3$ (PMT) a close
relation between the temperature behavior of (i) the diffuse scattering, (ii) the
dielectric response and (iii) the amplitude of the displacements of Pb ions from the special
Wyckoff positions in the perovskite structure~\cite{jpcm17,epjb40} has been found. Specifically, 
it was observed that the
susceptibility of the dynamic (quasi-elastic) component of the diffuse scattering in
PMN follows well the peak of the dielectric permittivity~\cite{jpcm17} and that
the intensity of the diffuse scattering in PMT matches closely
the amplitude of the Pb displacements~\cite{epjb40}.

Most of the studies on relaxor ferroelectrics have concentrated on
temperature and electric-field effects. Recently, interest
in the properties of the relaxors under hydrostatic pressure has
aroused~\cite{samara1996,yasuda1997,nomura1999,samara2000,chaabane2001,nawrocik2003,
chaabane2003}. Hydrostatic pressure applied to relaxors usually causes:
(i) the suppression of the peak in the dielectric permittivity,
as found, {\it e.g.}, in PMN~\cite{nawrocik2003} or in
PbIn$_{1/3}$Nb$_{1/2}$O$_3$ (PIN)~\cite{yasuda1997} and
(ii) the decrease in the intensity of the diffuse scattering as observed also, {\it e.g.},
in PMN~\cite{chaabane2003} and PIN~\cite{nomura1999}. However, up to now there exists no
attempt to link these two effects to the underlying structural changes in these systems.

Here we present a detailed structural investigation under
pressure, aimed to clarify the microscopic origin for the anomalous
pressure effects reported in relaxors.
We applied neutron diffraction taking advantage of a better sensitivity on the oxygen position 
as compared to X-ray studies.
The experiments were carried out on PMT, which at ambient pressure, shows a broad frequency-dependent
anomaly of the dielectric response with a maximum at a frequency of 10 kHz in the vicinity of 
170 K~\cite{smol1}.
In the entire range of temperatures and applied electric fields, the symmetry of PMT is known
to remain cubic (space group Pm$\bar{3}$m)~\cite{smol1,epjb40,lu}.

The powder diffraction experiment was carried out at ambient temperature on the
multi-detector high-resolution powder diffractometer HRPT~\cite{hrpt}
at the spallation neutron source SINQ ~\cite{sinq}
(Switzerland). Measurements
were performed with a neutron
wavelength of $\lambda=1.494 \rm~\AA$. Typical
exposure times were $\sim15$ hours.
The polycrystalline sample of PMT had a
volume of $\le100$~mm$^3$ and was loaded into an encapsulated zero-matrix TiZr
gasket \cite{Marshall} with a deuterated methanol-ethanol (4:1) mix as
a pressure-transmitting medium thus enabling hydrostatic pressure conditions up to the
highest investigated pressure of $\sim$$7$~GPa.
To apply pressure, the VX3 version of the Paris-Edinburgh press~\cite{pe2} was used.
A detailed description of this opposed-anvil set-up is given in
Ref.~\cite{aplhrpt}. The values of the applied pressure were estimated to better than $\pm 0.2$~GPa from the
change in the unit cell volume of PMT. To this end the third-order Birch-Murnaghan
equation of state~\cite{birch} was used with parameters taken from
Ref.~\cite{chaabane2003}.
%The crystal structure refinement was performed using
%Fullprof~\cite{rodriges}. % with the use of its internal table for neutron scattering lengths.
%
%%%%%%%%%%%%%%%%%%%%%%%%%%%%%%%%%%%%%%%%%%%%%%%%%%%%%%%%%%%%%%%%%%%%%%%%%%%%%%%%%%%%%%%%%
\begin{figure}[h]
%  \centering
   \includegraphics[width=0.5\textwidth, angle=0]{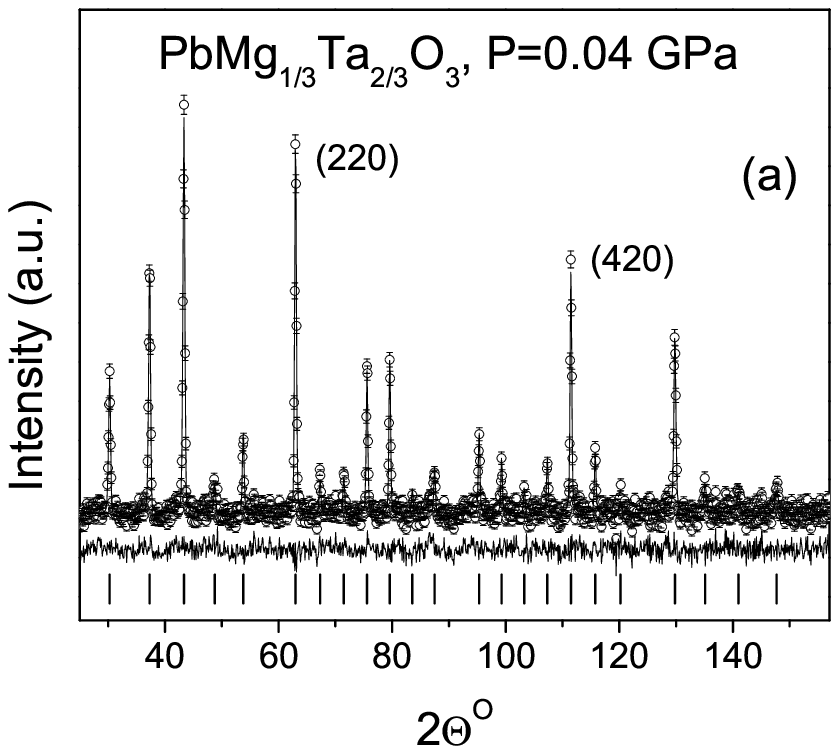}
   \includegraphics[width=0.5\textwidth, angle=0]{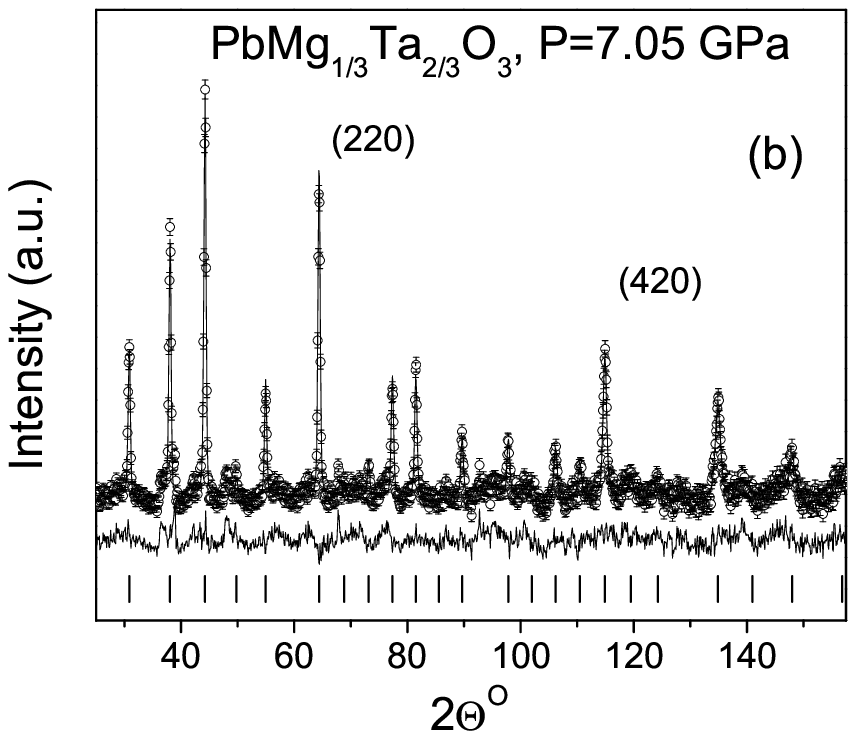}
   \caption{Neutron powder diffraction patterns of PMT collected at nominal pressures of
            (a) P=0.04 GPa and (b) P=7.05 GPa. Observed data points with background 
            originating from the pressure cell subtracted,
            calculated profiles and difference curves are shown.
            The row of ticks corresponds
            to the calculated positions of diffraction peaks. The peak
            in the vicinity of $2\rm\theta=45^o$ which is not fitted well within
            our model is due to scattering from a impurity of pyrochlore~\cite{epjb40}.}
\label{fig1}
\end{figure}
%%%%%%%%%%%%%%%%%%%%%%%%%%%%%%%%%%%%%%%%%%%%%%%%%%%%%%%%%%%%%%%%%%%%%%%%%%%%%%%%%%%%%%%%%
%

The crystal structure was refined with Fullprof~\cite{rodriges}
following a structural model of PMT developed in Ref.~\cite{epjb40}.
This study established that the structure is best described by Pb
ions displaced along the $<1~1~0>$ direction from the (0~0~0)
special position. The temperature factors of all ions are assumed to
be isotropic with the exception of the oxygen ions where anisotropic
components are considered too. For the Mg/Ta ions we assume a random
occupation of the Mg/Ta ions over the B-sites (0.5~0.5~0.5) of the
PMT perovskite structure according to the stoichiometric ratio. To
avoid unwanted correlation between the refined values of the
displacement amplitudes and the B- factor of Pb, latter was fixed to
a value of 1.14~$\rm \AA^2$ obtained at T = 300~K~\cite{epjb40}. We
note that we cannot rule out a displacement of the Mg/Ta or oxygen
ions but assert these displacements to be small in comparison to the
temperature factors~\cite{epjb40}.

%
%%%%%%%%%%%%%%%%%%%%%%%%%%%%%%%%%%%%%%%%%%%%%%%%%%%%%%%%%%%%%%%%%%%%%%%%%%%%%%%%%%%%%%%%%
\begin{figure}[h]
% \centering
  \includegraphics[width=0.35\textwidth, angle=0]{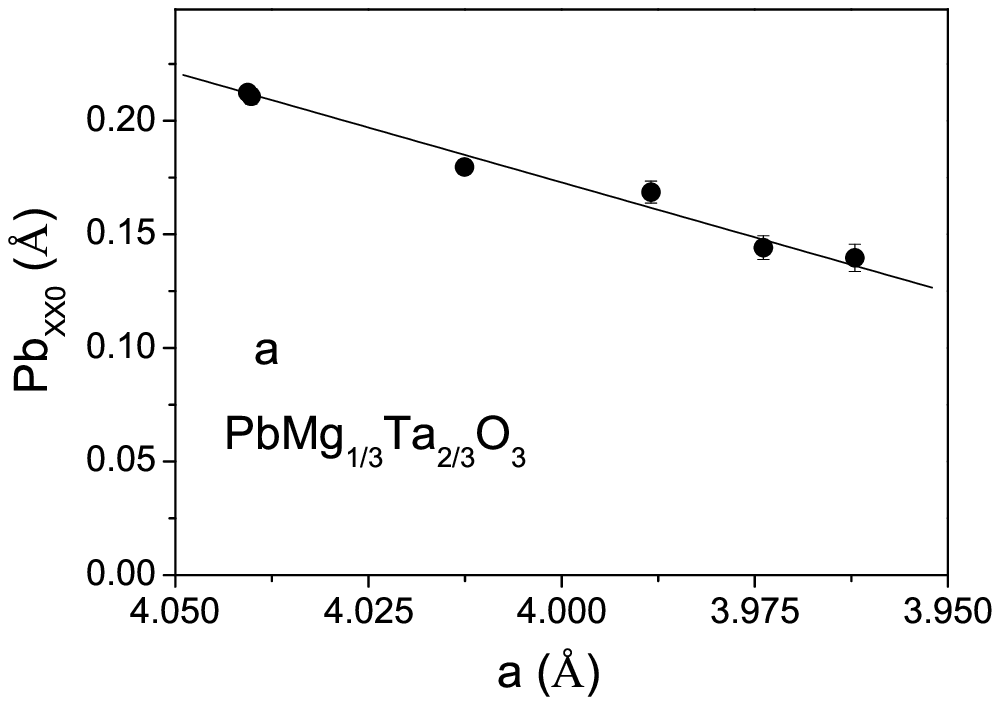}
  \includegraphics[width=0.35\textwidth, angle=0]{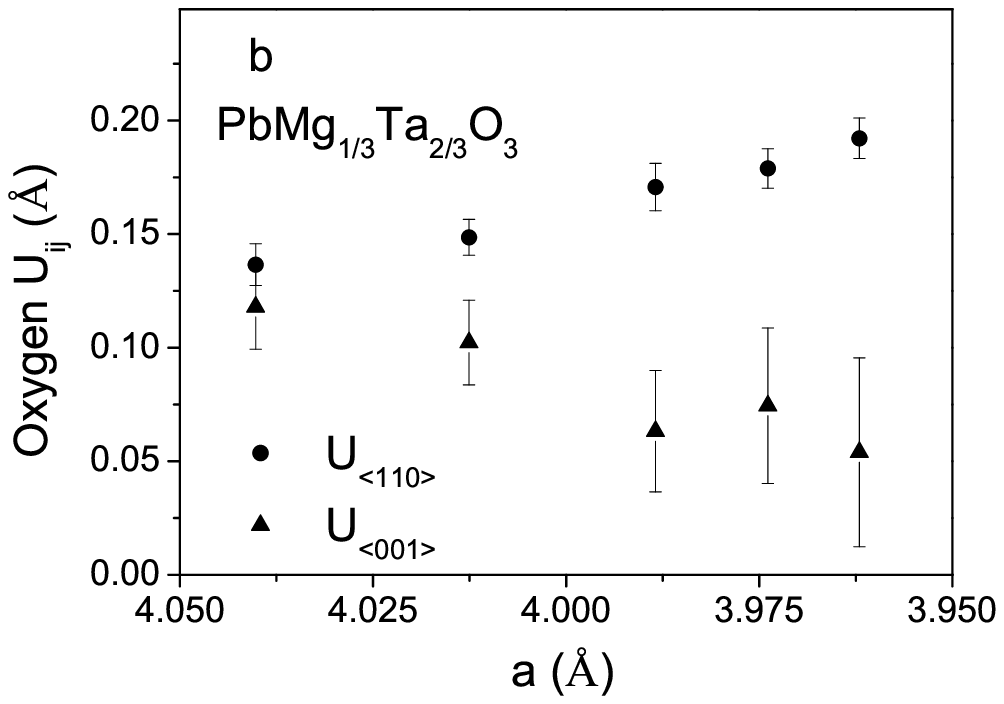}
  \includegraphics[width=0.35\textwidth, angle=0]{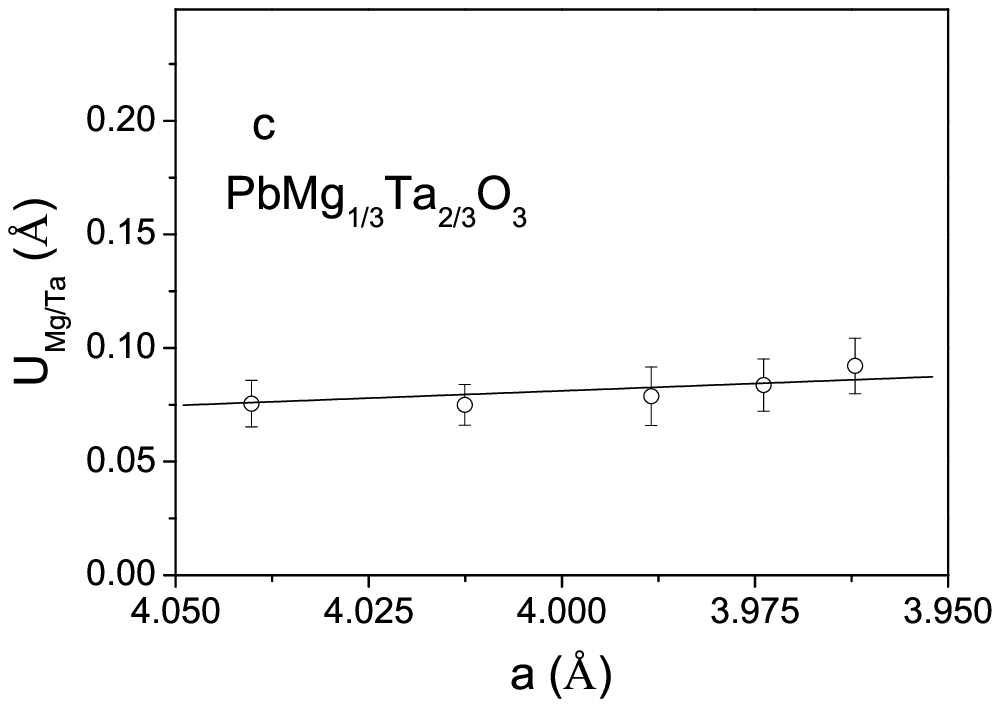}
  \caption{(a) Dependence of the amplitude of lead displacements Pb$\rm_{<XX0>}$ in PMT
           against lattice parameter. (b) Root mean square displacements of the oxygen ions
           against lattice parameter: note that the $<110>$ ($<001>$) direction points toward Pb (Mg/Ta).
           (c) Root mean square displacements of the Mg and Ta ions
           against lattice parameter. All graphs have the same scale to emphasize the
           significant difference in the pressure dependences of these parameters.
           Values of corresponding pressures can be extracted from Table~\ref{tab1}.}
\label{fig2}
\end{figure}
%%%%%%%%%%%%%%%%%%%%%%%%%%%%%%%%%%%%%%%%%%%%%%%%%%%%%%%%%%%%%%%%%%%%%%%%%%%%%%%%%%%%%%%%%
%

Figure~\ref{fig1} shows representative observed and calculated diffraction patterns obtained for
PMT close to ambient and at the highest pressure of nominally $P=7.05$~GPa. Apart from the trivial shift of all
peaks to higher scattering angles related to the contraction of the PMT lattice,
the effect of pressure is to modify significantly the relative peak intensities, as seen
for example on the (220) and (420) reflection. Latter effect results from an isostructural displacements of
the ions within the unit cell and which may be rationalized by a Rietveld analysis of the data.

Figure~\ref{fig2} shows the evolution of the lead displacements
(Pb$\rm_{<XX0>}$) and the root mean-square (RMS) displacements of
Mg/Ta and oxygen in PMT as a function of the lattice
parameter~\cite{comment1}. These quantities vary approximately
linearly with the lattice parameter, however, with opposite slopes
and with different rates: e.g., whereas the displacement of
Pb$\rm_{<XX0>}$ {\it decreases} by about $0.07 \rm\AA$ as the
lattice contracts, the RMS of the Mg/Ta ions {\it increases} by
about $0.02 \rm\AA$. It is worth noting that all observed changes in
the Pb$\rm_{<XX0>}$ displacements and
 RMS displacements of the other ions (e.g. Mg/Ta, 20\%) are considerably larger than that of the cell
volume of PMT ($\sim6\%$).

%
%%%%%%%%%%%%%%%%%%%%%%%%%%%%%%%%%%%%%%%%%%%%%%%%%%%%%%%%%%%%%%%%%%%%%%%%%%%%%
\begin{table}
\caption{Pressure, lattice parameter and
amplitude of lead displacement of PMT (equation of state with $B=104$~GPa, $B'=4.7$ taken from~\cite{chaabane2003}).
}
\label{tab1}
\begin{tabular}{|c|c|c|}
\hline
Pressure (GPa) & Lattice parameter ($\rm\AA$) & Pb displacement ($\rm\AA$)\\
\hline\hline
0.04 & 4.04019(7)  & 0.211(3) \\
2.28 & 4.01260(7)  & 0.180(3) \\
4.47 & 3.98822(9)  & 0.169(5) \\
5.86 & 3.9738(1)   & 0.144(5) \\
7.05 & 3.9620(2)   & 0.140(6) \\
\hline
\end{tabular}
\end{table}
%%%%%%%%%%%%%%%%%%%%%%%%%%%%%%%%%%%%%%%%%%%%%%%%%%%%%%%%%%%%%%%%%%%%%%%%%%%%%%

\begin{figure}[h]
% \centering
  \includegraphics[width=0.4\textwidth, angle=0]{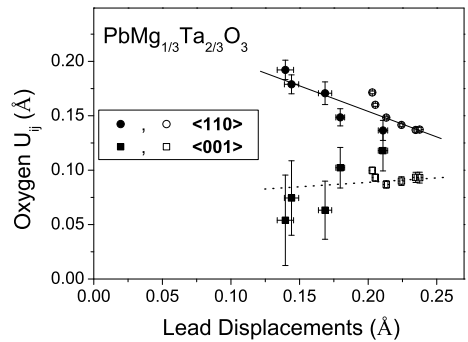}
  \caption{Components of the oxygen thermal ellipsoid versus the amplitude
           of lead displacements. Data points shown by open symbols are taken from earlier
           temperature-dependent studies of the structure of PMT~\cite{epjb40} at
           ambient pressure. These points were measured at
           T=588, 450, 300, 200, 85 and 1~K correspondingly. We
           note that smaller Pb displacements correspond to higher temperatures.}
\label{fig3}
\end{figure}

Further inspection of Fig.~\ref{fig2} shows that the decrease in the Pb$\rm_{<XX0>}$
is accompanied by an increase in the anisotropy of the oxygen
temperature factor. However, a reliable analysis of the components of the anisotropic
temperature factors requires high statistical quality of the data. This is hardly
achieved in the present experiment under high pressures and correspondingly small
sample volumes. Thus, in order to analyze
unambiguously the change in the anisotropy of the oxygen vibrations we have combined
earlier temperature dependent studies of the PMT structure~\cite{epjb40} with the
present results. It turns out that the amplitude of the Pb$\rm_{<XX0>}$ is a suitable
variable common to both the pressure- and the temperature-dependent studies.
Figure~\ref{fig3} thus shows the components of the oxygen thermal ellipsoid in relation to the
amplitude of Pb displacements. It is clearly seen that the $<110>$ and $<001>$ components
of the oxygen temperature factor evolve in the opposite direction: whereas the component directed along
the Pb-O bond increases significantly as the amplitude of the Pb displacements decreases,
the component directed along the Mg/Ta-O bond exhibits a slight decrease.
In fact, the latter component of the oxygen temperature factor is the only structural
parameter which has a similar relative change as the PMT cell volume.

The above evidenced pressure-induced changes of the
temperature factors in PMT are by far anomalous and may be contrasted to what is expected
for a simple solid. In the latter case
the temperature factors are related with the phonons and thus should
decrease with increasing pressure at a rate of approximately the average
Gr$\rm \ddot u$neisen-parameter, that is close to 1. This implies the relative
change of the temperature factors to scale with the change in the unit cell volume.
Obviously, the corresponding changes in PMT are in excess to such scaling and even show opposite
trends for Mg/Ta and for one of the components of the oxygen temperature factors.

%
%%%%%%%%%%%%%%%%%%%%%%%%%%%%%%%%%%%%%%%%%%%%%%%%%%%%%%%%%%%%%%%%%%%%%%%%%%%%%%%%%%%%%%%%%
%%%%%%%%%%%%%%%%%%%%%%%%%%%%%%%%%%%%%%%%%%%%%%%%%%%%%%%%%%%%%%%%%%%%%%%%%%%%%%%%%%%%%%%%%
%

Due to the fact that the perovskite-type materials constitute a significant part of the
earth mantle, their properties under high pressure are widely studied.
However, we are not aware that substantial changes of the atomic temperature
factors in these compounds have been reported before.  E.g., the temperature factors
of the ions in GdAlO$_3$ and GdFeO$_3$ crystals do not change significantly up to pressures
of about 8~GPa~\cite{ross1}. The relative changes of the atomic positions in these materials
all scale well with the changes in the unit cell volume.

On the other hand, it is known that hydrostatic pressure suppresses the ferroelectric
properties in perovskites~\cite{basset}. For example, at sufficiently high pressures
BaTiO$_3$~\cite{BaTiO3press} and KNbO$_3$~\cite{KNbO3press} restore the paraelectric
cubic structure~\cite{comment2}. However, even such significant changes
are not accompanied by profound changes in the temperature factors. In the case of the
tetragonal-to-cubic pressure-induced phase transition of BaTiO$_3$ there is a shift
of the Ti ion by $\rm \sim0.02\AA$ and the temperature factor of Ti increases by a factor of 2,
whereas the structure parameters of the other ions remain essentially unchanged~\cite{BaTiO3press} .

From the discussion above we conclude that the pressure-induced
changes in the structural parameters of PMT are a peculiarity of
relaxors. The most significant of these changes are the decrease of
the amplitudes of Pb displacements and the increase of the component
of the anisotropic temperature factor of oxygen ions directed toward
Pb. On the other hand, the change in the temperature factor of Mg/Ta
is pronounced much less. From the analysis of the Bragg peak
intensities we cannot infer direct information about the short-range
correlations between the displaced ions. However, it is reasonable
to expect that as the value of the ionic displacements decreases,
the associated short-range order diminishes, which in turn causes
(i) the reduction of the diffuse scattering and (ii) the suppression
of the dielectric permittivity peak. Thus, our observations give a
microscopic explanation for the fore-mentioned suppressions in
lead-containing relaxors, namely by the anomalous correlated
displacements of the lead and oxygen ions. A quantitative model to
describe such relationship is a matter of future theoretical and
experimental studies. High pressure measurements of the diffuse
scattering using single crystals would be highly desirable.

\begin{acknowledgments}
We thank D. Francis (ISIS Facility) for providing us with TiZr gaskets.
This work was performed at the spallation neutron source SINQ,
Paul Scherrer Institut, Villigen (Switzerland) and was partially
supported by RFBR grant 05-02-17822 and by Grant of
President RF ss-1415.2003.2.
\end{acknowledgments}

%
% ****** End of file template.aps ******


\begin{references}
\bibitem{comment1} The decrease of the lattice parameter is an alternative variable to present the
increase of hydrostatic pressure.
\end{references}
\end{document}